\documentclass[structabstract]{aa}
\usepackage{graphicx}
\usepackage{natbib}
\usepackage{amsmath,amssymb,graphicx}
\begin{document}
%
%\linenumbers

\title{Rotation periods and astrometric motions of the Luhman\,16AB brown dwarfs by high-resolution lucky-imaging
monitoring\thanks{Based on data collected by MiNDSTEp with the Danish 1.54\,m telescope at the ESO La Silla Observatory.}}
   \subtitle{}
\titlerunning{Lucky-imaging monitoring of Luhman\,16AB}

   \author{
          L. Mancini \inst{1,2}
          \and
          P. Giacobbe\inst{2}
          \and
          S.~P. Littlefair\inst{3}
          \and
          J. Southworth\inst{4}
          \and
          V. Bozza\inst{5,6}
          \and
          M. Damasso\inst{2}
          \and
          M. Dominik\inst{7}
          \and
          M. Hundertmark\inst{7}
          \and
          U.~G. J{\o}rgensen\inst{8}
          \and
          D. Juncher\inst{8}
          \and
          A. Popovas\inst{8}
          \and
          M. Rabus\inst{9,1}
          \and
          S. Rahvar\inst{10}
          \and
          R.~W. Schmidt\inst{11}
          \and
          J. Skottfelt\inst{12,8}
          \and
          C. Snodgrass\inst{13}
          \and
          A. Sozzetti\inst{2}
          \and
          K. Alsubai\inst{14}
          \and
          D.~M. Bramich\inst{14}
          \and
          S. Calchi Novati\inst{15,5,16}
          \and
          S. Ciceri\inst{1}
          \and
          G. D'Ago\inst{16,5,6}
          \and
          R. Figuera Jaimes\inst{7,17}
          \and
          P. Galianni\inst{7}
          \and
          S.-H. Gu\inst{18,19}
          \and
          K. Harps{\o}e\inst{8}
          \and
          T. Haugb{\o}lle\inst{8}
          \and
          Th. Henning\inst{1}
          \and
          T.~C. Hinse\inst{20}
          \and
          N. Kains\inst{21}
          \and
          H. Korhonen\inst{22,8}
          \and
          G. Scarpetta\inst{5,6,16}
          \and
          D. Starkey\inst{7}
          \and
          J. Surdej\inst{23}
          \and
          X.-B. Wang\inst{18,19}
          \and
          O. Wertz\inst{23}
          }
          {
    % 1
    \institute{
    Max Planck Institute for Astronomy, K\"{o}nigstuhl 17, 69117 -- Heidelberg, Germany \\
    \email{mancini@mpia.de}
        \and
    % 2
    INAF -- Osservatorio Astrofisico di Torino, via Osservatorio 20, 10025 -- Pino Torinese, Italy
        \and
    % 3
    Department of Physics and Astronomy, University of Sheffield, Sheffield S3 7RH, UK
         \and
    % 4
    Astrophysics Group, Keele University, Keele ST5 5BG, UK
         \and
    % 5
    Department of Physics, University of Salerno, Via Giovanni Paolo II 132, 84084 -- Fisciano (SA), Italy
         \and
     % 6
     Istituto Nazionale di Fisica Nucleare, Sezione di Napoli, I-80126 Napoli, Italy
         \and
     % 7
     SUPA, University of St Andrews, School of Physics \& Astronomy, North Haugh, St Andrews, Fife KY16 9SS, UK
        \and
     % 8
     Niels Bohr Institute \& Centre for Star and Planet Formation, University of Copenhagen, {\O}stervoldgade 5, 1350 -- Copenhagen K, Denmark
          \and
     % 9
Instituto de Astrof\'{i}sica, Pontificia Universidad Cat\'{o}lica de Chile, Av. Vicu\~{n}a Mackenna 4860, 7820436 -- Macul, Santiago, Chile %
          \and
     % 10
Department of Physics, Sharif University of Technology, PO Box 11155-9161 Tehran, Iran
          \and
     % 11
Astronomisches Rechen-Institut, Zentrum f\"{u}r Astronomie, Universit\"{a}t Heidelberg, M\"{o}nchhofstrasse 12-14, 69120 -- Heidelberg, Germany
           \and
      % 12
Centre for Electronic Imaging, Dept. of Physical Sciences, The Open University, Milton Keynes MK7 6AA, UK
           \and
      % 13
Planetary and Space Sciences, Dept. of Physical Sciences, The Open University, Milton Keynes MK7 6AA, UK %
           \and
      % 14
Qatar Environment and Energy Research Institute, Qatar Foundation, Tornado Tower, Floor 19, PO Box 5825, Doha, Qatar
           \and
      % 15
%Institute for Astronomy, University of Edinburgh, Blackford Hill, Edinburgh EH9 3HJ, UK
%           \and
      % 15
NASA Exoplanet Science Institute, MS 100-22, California Institute of Technology, Pasadena, CA 91125, USA
           \and
      % 16
Istituto Internazionale per gli Alti Studi Scientifici (IIASS), 84019 -- Vietri Sul Mare (SA), Italy
           \and
      % 17
European Southern Observatory, Karl-Schwarzschild-Strasse 2, 85748 -- Garching bei M\"{u}nchen, Germany
           \and
      % 18
Yunnan Observatories, Chinese Academy of Sciences, Kunming 650011, China
           \and
      % 19
Key Laboratory for the Structure and Evolution of Celestial Objects, Chinese Academy of Sciences, Kunming 650011, China
            \and
      % 20
Korea Astronomy and Space Science Institute, Daejeon 305-348, Republic of Korea
             \and
      % 21
Space Telescope Science Institute, 3700 San Martin Drive, Baltimore, MD 21218, USA
             \and
      % 22
Finnish Centre for Astronomy with ESO (FINCA), University of Turku, V\"{a}is\"{a}l\"{a}ntie 20, 21500 -- Piikki\"{o}, Finland
             \and
      % 23
Institut d'Astrophysique et de G\'{e}ophysique, Universit\'{e} de Li\`{e}ge, 4000 -- Li\`{e}ge, Belgium
}
%   \date{Received ; Accepted}

\abstract
% % 6 {} token are mandatory
{Photometric monitoring of the variability of brown dwarfs can provide useful information about the structure of clouds in their cold atmospheres.
The brown-dwarf binary system Luhman\,16AB is an interesting target for such a study, as its components stand at the L/T transition and show high levels of variability. Luhman\,16AB is also the third closest system to the Solar system, allowing precise astrometric investigations with ground-based facilities.}
%  % aims heading (mandatory)
{The aim of the work is to estimate the rotation period and study the astrometric motion of both components.}
%% methods heading (mandatory)
{We have monitored Luhman\,16AB over a period of two years with the lucky-imaging camera mounted on the Danish 1.54\,m telescope at La Silla, through a special $i+z$ long-pass filter, which allowed us to clearly resolve the two brown dwarfs into single objects.
An intense monitoring of the target was also performed over 16 nights, in which we observed a peak-to-peak variability of $0.20 \pm 0.02$\,mag and $0.34 \pm 0.02$\,mag for Luhman\,16A and 16B, respectively.}
%% results heading (mandatory)
{We used the 16-night time-series data to estimate the rotation period of the two components. We found that Luhman\,16B rotates with a period of $5.1 \pm 0.1$\,hr, in very good agreement with previous measurements. For Luhman\,16A, we report that it rotates slower than its companion and, even though we were not able to get a robust determination, our data indicate a rotation period of roughly 8\,hr. This implies that the rotation axes of the two components are well aligned and suggests a scenario in which the two objects underwent the same accretion process. The 2-year complete dataset was used to study the astrometric motion of Luhman\,16AB. We predict a motion of the system that is not consistent with a previous estimate based on two months of monitoring, but cannot confirm or refute the presence of additional planetary-mass bodies in the system.}
%% conclusions heading (optional), leave it empty if necessary
{}


\maketitle

% Sect. 1
%%%%%%%%%%%%%%%%%%%%%%%%%%%%%%%%%%%%%%%%%%%%%%%%%%%%%%
\section{Introduction}
\label{sec_1}
%%%%%%%%%%%%%%%%%%%%%%%%%%%%%%%%%%%%%%%%%%%%%%%%%%%%%%
%
\begin{figure}
\centering
\includegraphics[width=9.0cm]{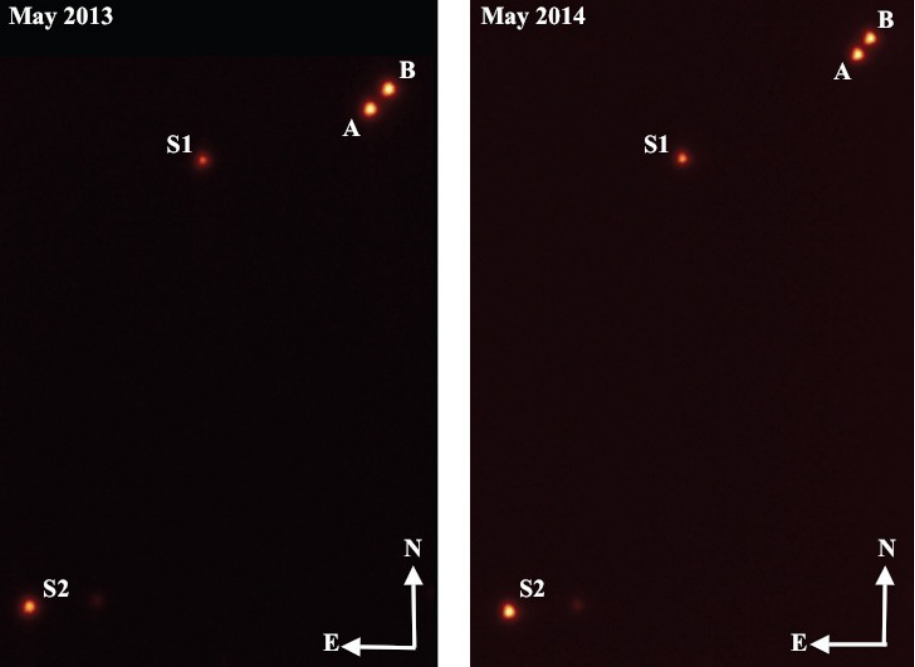}
\caption{Two images of the brown-dwarf binary system Luhman\,16AB, obtained with the LI camera mounted on the Danish 1.54\,m Telescope. The images were taken one year apart. The same two comparison stars are also present in the FOV.}%
\label{image}
\end{figure}

Brown dwarfs are very intriguing astrophysical objects due to their mass range, between the gas planets and the lightest M-type stars. They are classified into M$\rightarrow$L$\rightarrow$T$\rightarrow$Y classes, based on their spectral characteristics. Since brown dwarfs are not massive enough to trigger sufficient nuclear fusion reaction rates in their cores to sustain hydrostatic equilibrium, they are destined to gradually cool and this classification also represents an evolutionary sequence (e.g., \citealp{kirkpatrick:1999,burgasser:2006,cushing:2011,kirkpatrick:2012}).
As a brown dwarf cools, the temperature and gas pressure of its photosphere become such that it starts to be progressively dominated by molecular gas. When an L-type brown dwarf cools down to hot-Jupiter temperatures, the molecular gas can condense into thin or thick clouds or a mixture of them. Aerosols can form in its atmosphere through the condensation of very refractory species like metallic oxides, silicates, and iron 
\citep{lodders:1999,marley:2002,woitke:2003,woitke:2004,burrows:2006,helling:2008}. At the L$\rightarrow$T class transition ($T_{\mathrm{eff}} \approx 1200-1300$\,K), these clouds could break up and unveil the naked photosphere, causing a brown dwarf to be very variable over short timescales ($\sim$\,hr), depending on the inclination angle of the rotation axis \citep{ackerman:2001,burgasser:2002,marley:2010}.
It is also possible that the clouds can become more and more thin and eventually disappear, due to the increasing size of their particles and subsequent rain-out \citep{tsuji:2003,knapp:2004}.
Photometric time-variability monitoring of brown dwarfs is therefore a useful diagnostic tool to investigate the properties of clouds in their atmospheres, and periodic or quasi-periodic rapid variability has been found in several cases {both at optical and near-infrared wavelengths} (e.g., \citealp{artigau:2009,radigan:2012,buenzli:2012,heinze:2013,biller:2013,radigan:2014,buenzli:2014}).

In this context, the binary system Luhman\,16AB (aka WISE J104915.57$-$531906.1; \citealp{luhman:2013}), being composed of L7.5 (A component) and T0.5 (B component) brown dwarfs \citep{kniazev:2013,burgasser:2013}, represents an emblematic case for studying the L/T transition states of brown dwarfs and investigating the disappearance of clouds

A first 12-day long monitoring of this system was performed by \citet{gillon:2013} with the TRAPPIST 60\,cm telescope through a $i+z$ filter. Even though they were not able to resolve the two components of the system, they found a large variability ($\sim 10\%$) and, attributing this variability to the cooler component, estimated a rotation period of $4.87 \pm 0.01$\,hr for Luhman\,16B.

\begin{table}
\caption{Summary of the data collected with the LI camera mounted on the Danish 1.54\,m telescope.} %
\label{tab:summary} %
\centering     %
\begin{tabular}{lccc}
\hline\hline\\[-6pt]
Date & $N_{\rm obs}$ & Rotation period & Astrometry \\
\hline\\[-6pt]
\multicolumn{4}{l}{\textbf{2013:}} \\
2013.05.02 & 3 & No & No \\
2013.05.04 & 9 & No & Yes \\
2013.05.05 & 20 & No & Yes \\
2013.05.06 & 33 & No & Yes \\
2013.05.07 & 38 & No & Yes \\
2013.05.08 & 36 & No & No \\
2013.05.09 & 26 & No & Yes \\
2013.05.10 & 38 & No & No \\
2013.05.11 & 37 & No & No \\
2013.05.12 & 28 & No & No \\
2013.05.13 & 10 & No & Yes \\
2013.05.14 & 31 & No & Yes \\
2013.05.15 & 19 & No & Yes \\
2013.05.16 & 24 & No & No \\ [6pt] %
\multicolumn{4}{l}{\textbf{2014:}} \\
2014.04.19 & 3 & Yes & Yes \\
2014.04.20 & 26 & Yes & No \\
2014.04.21 & 2 & Yes & Yes \\
2014.04.22 & 42 & Yes & Yes \\
2014.04.23 & 53 & Yes & Yes \\
2014.04.24 & 66 & Yes & Yes \\
2014.04.25 & 64 & Yes & Yes \\
2014.04.26 & 39 & Yes & No \\
2014.04.27 & 40 & Yes & Yes \\
2014.04.28 & 54 & Yes & Yes \\
2014.04.29 & 66 & Yes & No \\
2014.04.30 & 15 & Yes & Yes \\
2014.05.01 & 56 & Yes & Yes \\
2014.05.02 & 52 & Yes & No \\
2014.05.04 & 65 & Yes & Yes \\
2014.05.05 & 65 & Yes & No \\
2014.05.09 & 3 & No & Yes \\
2014.05.17 & 3 & No & Yes \\
2014.05.27 & 2 & No & No \\
2014.05.31 & 1 & No & No \\
2014.06.07 & 2 & No & No \\
2014.06.14 & 2 & No & No \\
2014.06.17 & 1 & No & No \\
2014.06.20 & 2 & No & No \\
2014.06.21 & 1 & No & No \\
2014.06.23 & 1 & No & No \\
2014.07.04 & 1 & No & No \\
2014.07.16 & 1 & No & No \\
\hline
\end{tabular}
\tablefoot{$N_{\rm obs}$ is the number of observations for each single night.
Column 3 indicates if the data were used for estimating the rotation periods of the components of Luhman\,16.
Column 4 indicates if the data were used for the astrometric study of the components of Luhman\,16.}
\end{table}

Using the GROND instrument \citep{greiner:2008} mounted on the MPG 2.2\,m telescope, the two components were resolved and monitored for 4\,hr in four optical and three near-infrared (NIR) passbands simultaneously by \citet{biller:2013}. They found that the B component shows variability also in the NIR bands and reported a low-amplitude intrinsic variability in the $i$ and $z$ bands for the A component. A subsequent, additional 7-day monitoring with the TRAPPIST telescope led to the finding of a rotation period of $5.05 \pm 0.10$\,hr for Luhman\,16B \citep{burgasser:2014}, which is consistent with the previous measurement.

A fascinating surface map of Luhman\,16B was deduced by \citet{crossfield:2014} by using the Doppler-imaging technique on high-resolution spectra taken with the CRIRES spectrograph \citep{kaufl:2004} at the ESO Very Large Telescope (VLT). The surface of Luhman\,16B appears to be structured into large bright and dark regions, reasonably recognised as patchy clouds. Monitoring of these clouds for several hours allowed to confirm the rotation period of $\sim 5$\,hr for Luhman\,16B. Other useful information was extracted from the CRIRES data, which show that CO and H$_2$O absorption features dominate the spectra of both the components. The projected rotational velocities were estimated to be $17.66 \pm 0.1$\,km\,s$^{-1}$  and $26.16 \pm 0.2$\,km\,s$^{-1}$ for Luhman\,16A and 16B, respectively. This implies that component B should have a rotational axis inclined by less than $\sim30^{\circ}$ to the plane of the sky, assuming that its radius is roughly $1\,R_{\mathrm{Jup}}$ \citep{crossfield:2014}.

Another peculiarity of Luhman\,16 is that, after Alpha\,Centari\,AB, it is the nearest known binary system to the Solar system, at $\sim 2$\,pc. This occurrence allows astrometric studies to be performed for the two components of the system, in order to detect their orbital motion and measure their parallax. This was done by \citet{boffin:2014}, who monitored Luhman\,16AB for a period of two months with the FORS2 instrument at the ESO-VLT, measuring a distance of $2.020 \pm 0.019$\,pc and a proper motion of about 2.8$^{\prime \prime}$/year. Moreover, they found that the relative orbital motion of the two objects is perturbed, suggesting the presence of a substellar companion around one of the two components. However, the existence of a substellar object in the system was not confirmed by \citet{melso:2015}, who, based on multi-epoch images from the {\it Spitzer} Space Telescope and adaptive optics images from the VLT, did not detect any new companion.

In this work we present new long-term, high-resolution, photometric monitoring of the Luhman\,16 system over two years, performed through an optical broad-band filter (700-950\,nm) with a lucky-imaging camera. The new data have been used to study the variability in red-optical light (Sect.\,2) and revise the astrometric motion of the two brown dwarfs (Sect.\,3). Sect.\,4 summarises our results.

% Sect. 2
%%%%%%%%%%%%%%%%%%%%%%%%%%%%%%%%%%%%%%%%%%%%%%%%%%%%%%
\section{Photometry}
\label{photometry}
%%%%%%%%%%%%%%%%%%%%%%%%%%%%%%%%%%%%%%%%%%%%%%%%%%%%%%

%%%%%%%%%%%%%%%%%%%%%%%%%%%%%%%%%%%%%%%%%%%%%%%%%%%%%%
\subsection{Observations and data reduction}
\label{observations}
%%%%%%%%%%%%%%%%%%%%%%%%%%%%%%%%%%%%%%%%%%%%%%%%%%%%%%

%
\begin{figure*}
\centering
\includegraphics[width=18.0cm]{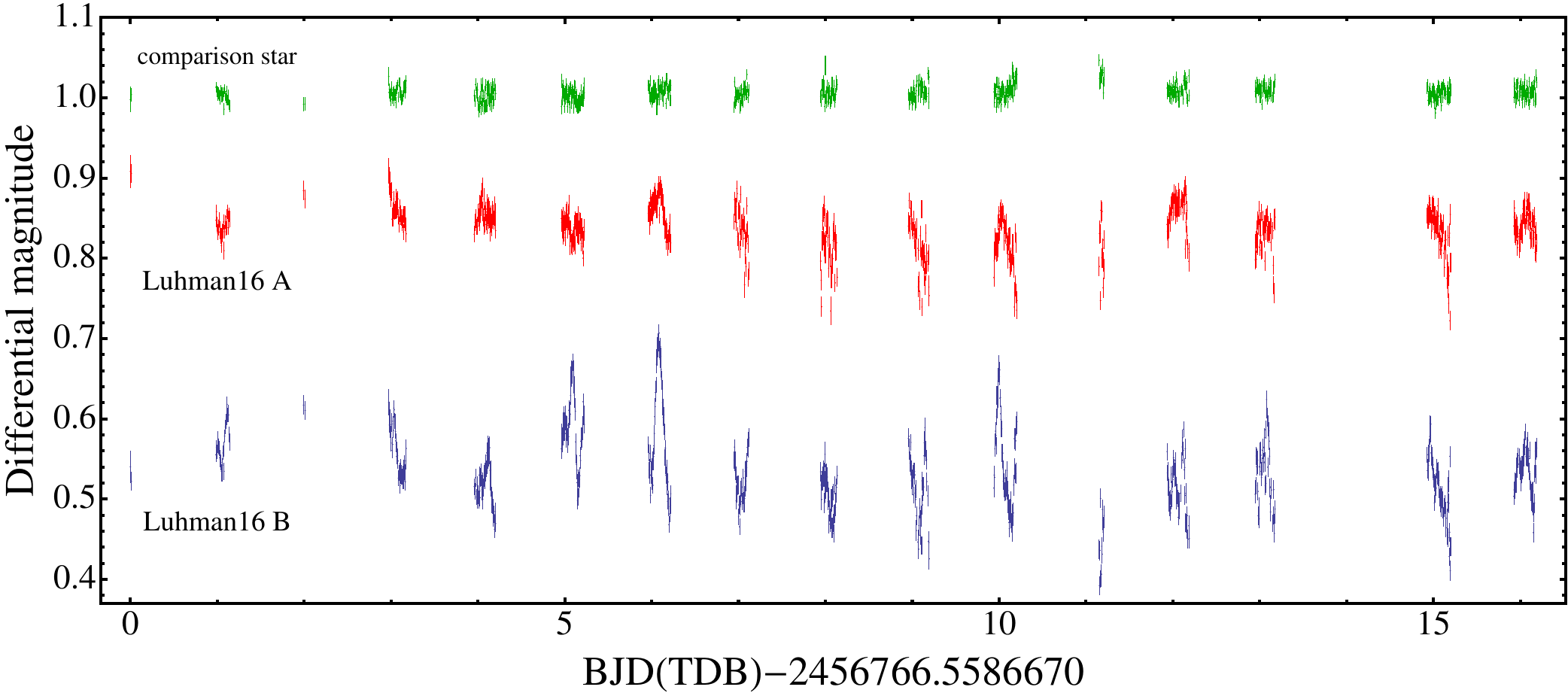}
\caption{Globally normalized unbinned PSF photometry for the two components of Luhman\,16 and for one of the
comparison stars used in the reduction (shifted along the $y$ axis), based on 16 nights of photometric monitoring
with the LI camera. Red points refer to Luhman\,16A, blue to Luhman\,16B and green to a comparison star.}%
\label{fotometria1}
\end{figure*}
We monitored the brown-dwarf binary system Luhman\,16AB in lucky-imaging mode with the EMCCD (electron-multiplying charge-couple device) instrument \citep{skottfelt:2015b} mounted on the Danish 1.54\,m Telescope, located at the ESO Observatory in La Silla. This device consists of an Andor Technology iXon+ model 897, $512\times512$ pixel EMCCD Lucky-Imaging (LI) Camera. Its pixel scale is 0.09\,arcsec\,pixel$^{-1}$, resulting in a field of view of $45\times45$\,arcsec$^{2}$. The LI camera is mounted behind a dichroic mirror, which acts as a long-pass filter roughly corresponding to a combination of the SDSS $i$ and $z$ filters \citep{skottfelt:2013,skottfelt:2015}.

The Luhman\,16AB system was monitored during two seasons. The first season started on May 2, 2013 and finished on May 16, 2013. 352 images were collected during 14 nights, with an exposure time of 3\,min. The second season started on April 19, 2014 and finished on July 16, 2014, and comprises 728 images with an exposure time of 5\,min. In particular, a dense monitoring of the target was performed between April 19 and May 5, 2014, in which we obtained 708 images spread over 16 nights, of which 14 were consecutive; after that, the target was observed few times at intervals of several nights. In total, considering both the seasons, we observed the target for 42 nights. However the quality of the data is not the same for all the nights. In particular, the two components were not well resolved in the images observed with a seeing $\gtrapprox 1$\,arcsec, and such data turned out to be of no use to our analysis.

Table\,\ref{tab:summary} summarises the data collection. Bias and dome flat-field frames were taken every night and used to calibrate the science images. A comparison between the 2013 and 2014 positions of the brown dwarfs, with respect to two comparison stars, is illustrated in Fig.\,\ref{image}. It also clearly shows that our observations consistently resolved the two components of the brown-dwarf binary.

Each observation consists of a data cube containing 1800 (2013 season) or 3000 (2014 season) single exposures of exposure time 0.1\,s.
%Master bias and (dome) flat-field images were created by median-combining a suitable number of frames, which were %taken at the beginning of each night of observations, and used to calibrate the scientific images. For this purpose
They were calibrated using algorithms described in \citet{harpsoe:2012} and the output of the reduction of a single observation was a ten-layer image cube fits file. The flux from each of the brown dwarfs was extracted through PSF photometry using a modified version of the {\sc defot} pipeline, written in IDL\footnote{The acronym IDL
stands for Interactive Data Language and is a trademark of ITT
Visual Information Solutions.} (\citealp{southworth:2014} and reference therein). An ensemble of comparison stars in the FOV were used for this purpose. All the data will be made available at the CDS.
The resulting light curves of both the components, corresponding to the 16-night dense monitoring, are shown in Fig.\,\ref{fotometria1}, together with that of a comparison star.
%They are also illustrated in the panels of Fig.\,\ref{fotometria2}, in which the photometric variations of the two components is %highlighted, resulting in a peak-to-peak variability of $0.20 \pm 0.02$\,mag for Luhman16\,A and $0.34 \pm 0.02$\,mag for %Luhman16\,B.
The peak-to-peak variability is $0.20 \pm 0.02$\,mag for Luhman16\,A and $0.34 \pm 0.02$\,mag for Luhman16\,B.
The two data sets were assembled in tables, which are available at the CDS, each containing the timestamps, differential magnitudes and corresponding uncertainties.

%%%%%%%%%%%%%%%%%%%%%%%%%%%%%%%%%%%%%%%%%%%%%%%%%%%%%%
\subsection{Time-series analysis}
\label{timeseries}
%%%%%%%%%%%%%%%%%%%%%%%%%%%%%%%%%%%%%%%%%%%%%%%%%%%%%%
%
%
\begin{figure*}
\centering
\includegraphics[width=18.0cm]{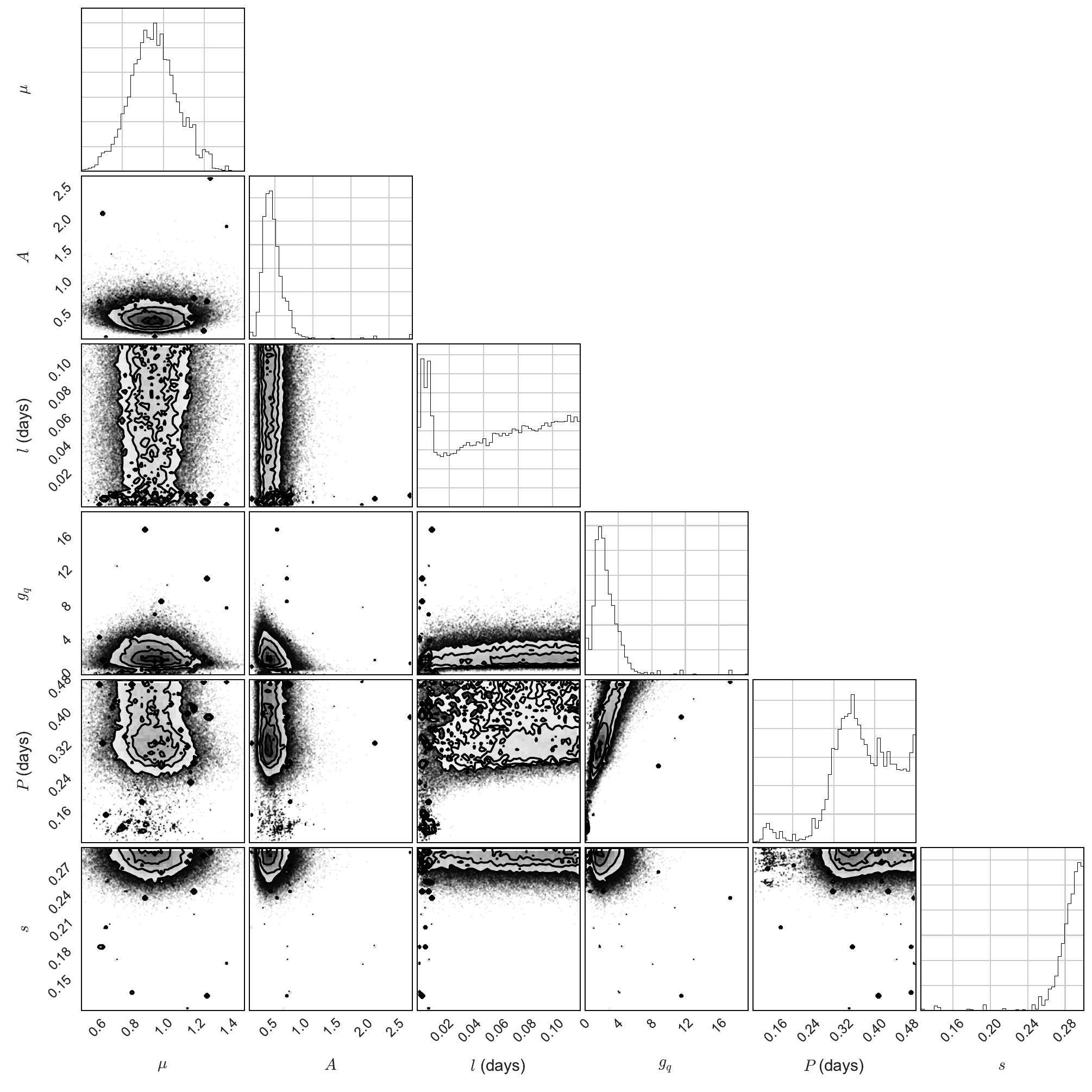}
\caption{Samples from the posterior probability distributions of the correlations for the six fitted
parameters (see text) of Luhman\,16A. $\mu$ is the mean level of the light curve. The full file of samples
from the posterior distributions is available on request by sending an email to the first author.}%
\label{cornerPlot_A}
\end{figure*}
\begin{figure*}
\centering
\includegraphics[width=18.0cm]{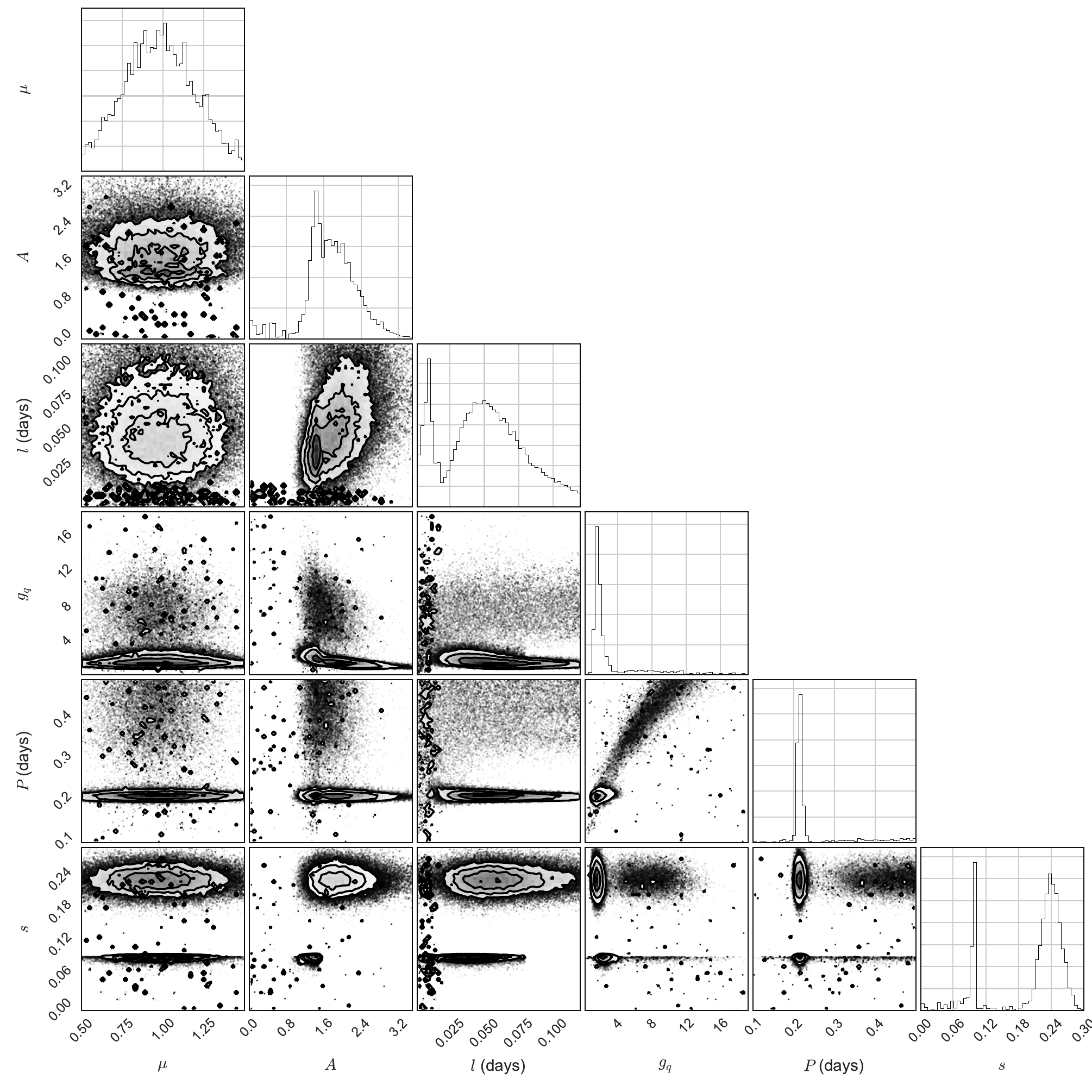}
\caption{Samples from the posterior probability distributions of correlations for the six fitted
parameters (see text) of Luhman\,16B. $\mu$ is the mean level of the light curve. The full file of samples
from the posterior distributions is available on request by sending an email to the first author.}%
\label{cornerPlot_B}
\end{figure*}
We analysed the 16-night photometric time series taken in 2014 (Fig.\,\ref{fotometria1}), to investigate if their variability is periodic and thus ascribable to the rotation of the brown dwarfs and caused by possible inhomogeneities (clouds) in their atmospheres. The monitoring performed in all other nights (see Table\,\ref{tab:summary}) is too sparsely sampled to yield robust results.

We attempted to measure the rotation periods of Luhman\,16AB by fitting the photometric data for each component individually with a Gaussian process model. The data points in the light curves were modelled as being drawn from a multivariate Gaussian distribution; time correlations between data points were reflected in non-zero values for the off-diagonal elements of the covariance matrix. Following \citet{vanderburg:2015} we parameterised the covariance matrix with a quasi-periodic kernel function. In principle this is an improvement on a periodogram analysis since the evolving cloud features on Luhman\,16AB produce variability which is neither sinusoidal nor strictly periodic. The kernel function adopted is given by
{\small
\begin{equation}
k_{ij} = A^{2} \exp \left[ \frac{-(x_{i}-x_{j})^{2}}{2l^{2}} \right] \exp \left[ \frac{ -\sin^{2} \left( \frac{\pi (x_{i}-x_{j})}{P} \right)}{g_{\mathrm{q}}^{2}} \right] + s^{2}\delta_{ij}, \nonumber
\end{equation}}
where $A$ is the amplitude of the correlation, $l$ is the timescale of the exponential decay term, $P$ is the rotation period, $g_{\mathrm{q}}$ is a scaling factor for the exponentiated sinusoidal term and $s$ is a white noise hyper-parameter. $g_{q}$ and $l$ are hyper-parameters related to the time-evolving features (clouds) in the atmospheres of the brown dwarfs: the first is a scale factor which takes into account the changes in sizes of the features, and the second is related to their lifetimes. Essentially, the scaling factor $g_{\mathrm{q}}$ affects the regularity of the resulting lightcurve, with smaller values producing models which are increasingly sinusoidal.

Our analysis made use of {\sc george} \citep{foreman:2014}, a Gaussian-process library that uses a fast matrix inversion method \citep{ambikasaran:2014}, to implement our Gaussian process. We explored the posterior distributions of the hyper-parameters using a Markov-Chain Monte-Carlo analysis with the affine invariant ensemble sampler within {\em emcee} \citep{foreman:2013}.

For Luhman\,16A, the posterior distribution for the rotation period $P$ is not well defined (Fig.\,\ref{cornerPlot_A}). A period of around 8 hours is preferred, but a wide range of rotation periods are compatible with the data. This is essentially caused by the fact that Luhman\,16A is rotating slower than its companion and that the evolutionary timescale of the global weather on Luhman\,16A is roughly one day, while the monitoring with the Danish Telescope lasted $4-5$\,hr per night. Nevertheless, it is interesting to remark that, based on comparison with $v\sin{i}$ data \citep{crossfield:2014}, a rotation period of $\sim 8$\,hr is exactly what is expected assuming that the rotational axes of the two components are aligned and likely implies that they experienced the same accretion process \citep{wheelwright:2011}. Such spin-orbit alignment has already been observed in the very low mass dwarf-binary regime \citep{harding:2013} and can be explained by different formation theories (see discussion in \citealp{harding:2013}).

For Luhman\,16B, the posterior distribution for the period is sharply peaked around 5 hours, with a `background' of low probability covering a wide range of periods (Fig.\,\ref{cornerPlot_B}). We modelled the samples from the posterior distribution with a Gaussian mixture model with two components. The sharp spike is well modelled by a Gaussian distribution, yielding an estimate of the rotation period for Luhman\,16B of $5.1\pm0.1$\,hr, which is fully consistent with the value of $5.05\pm0.10$\,hr estimated by \citet{burgasser:2014}. In Figs.\ \ref{bestFit_A} and \ref{bestFit_B} we show the light curves for Luhman\,16A and Luhman\,16B respectively, together with a representation of the best fit implied by the Gaussian process model.

%We used the Generalized
%Lomb-Scargle algorithm (GLS; \citealp{zechmeister:2009}) for this purpose. While for the A component we did not find any %significant peak, for the B component we found a peak in the periodogram at 4.9\,hr, in agreement with the rotation period found %by \citet{gillon:2013} and \citet{crossfield:2014}. However, the GLS returned a more significant peak at $6.2$\,hr, see Fig.\ref
%{gls}.

%
\begin{figure*}
%\centering
\includegraphics[trim=7cm 3cm 0cm 4cm, clip=true, totalheight=0.52\textheight, angle=0]{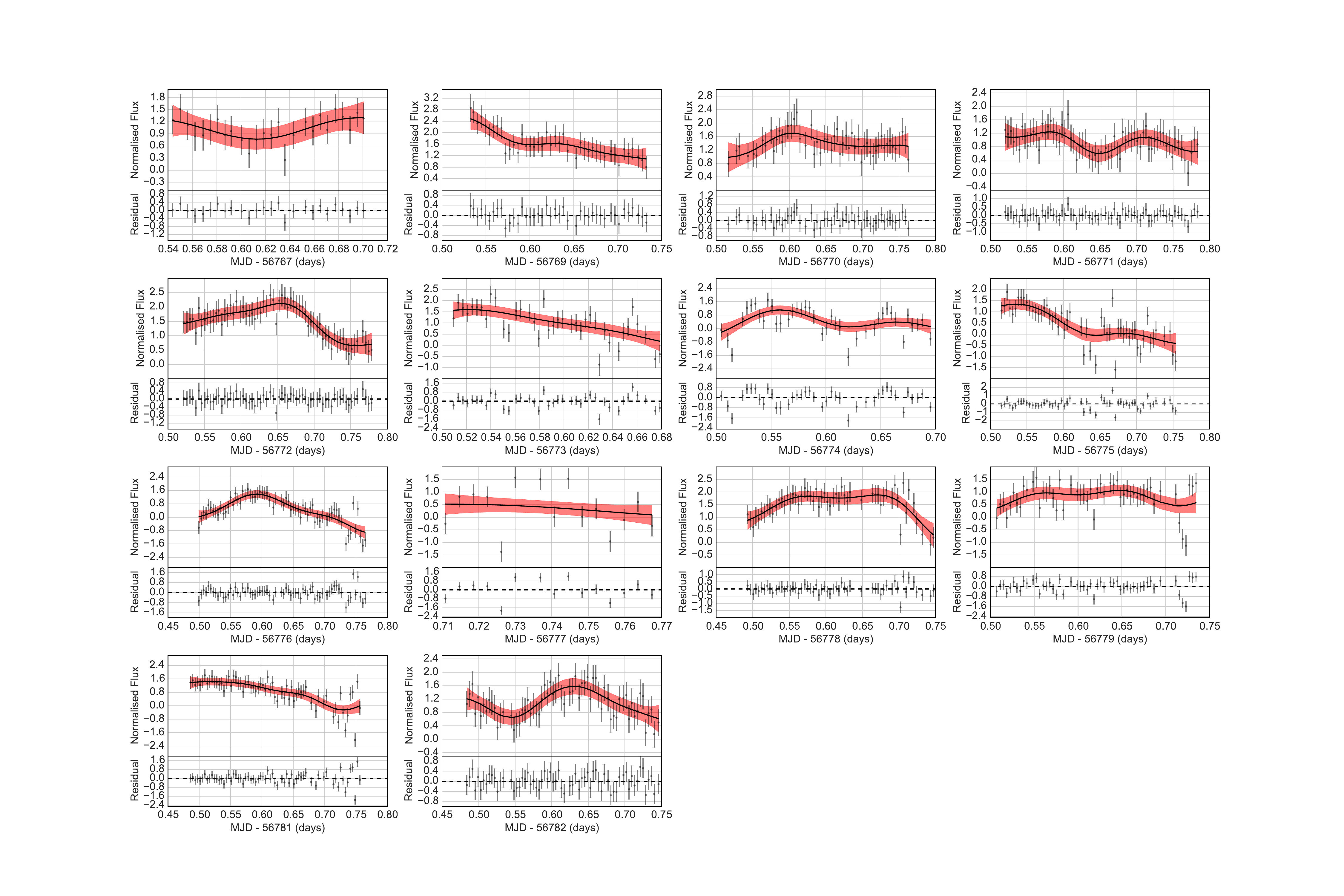}
\caption{Single-night light curves for Luhman\,16A, together with a representation of the best
fit implied by the Gaussian process model. The solid black line represents the mean of 300 samples
drawn from the Gaussian Process model conditioned on the full dataset. The red shaded region
represents the 2-$\sigma$ confidence interval, as estimated from these samples.}%
\label{bestFit_A}
\end{figure*}
\begin{figure*}
%\centering
\includegraphics[trim=7cm 3cm 0cm 4cm, clip=true, totalheight=0.52\textheight, angle=0]{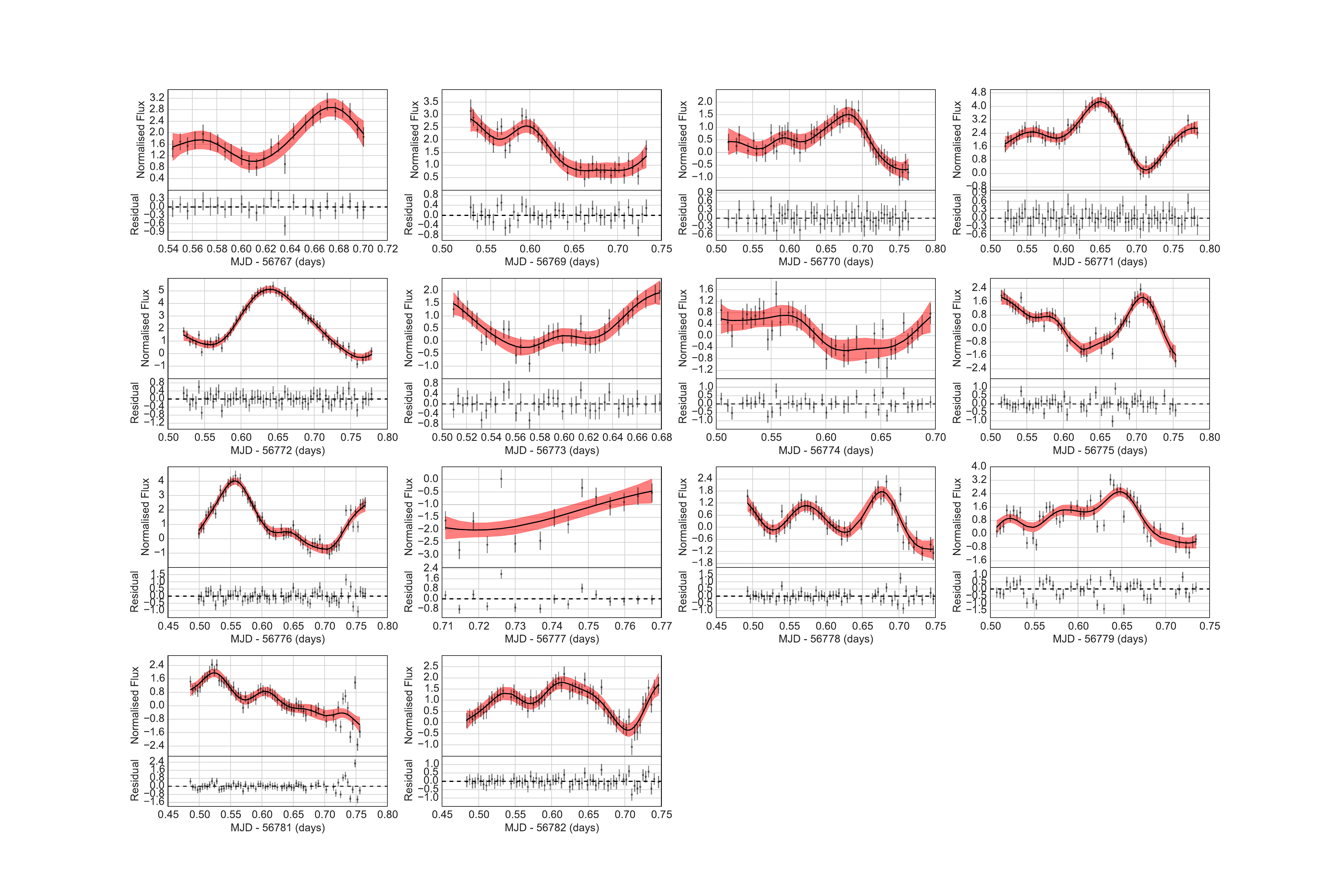}
\caption{Single-night light curves for Luhman\,16B, together with a representation of the best fit implied by the Gaussian process model.}%
\label{bestFit_B}
\end{figure*}
%

% Sect. 3
%%%%%%%%%%%%%%%%%%%%%%%%%%%%%%%%%%%%%%%%%%%%%%%%%%%%%%
\section{Astrometry}
\label{astrometry}
%%%%%%%%%%%%%%%%%%%%%%%%%%%%%%%%%%%%%%%%%%%%%%%%%%%%%%

\subsection{Astrometic reduction}
\label{astrometric_reduction}

For each image, we extracted the $x,y$ (plate) coordinates via PSF fitting (using SExtractor and PSFExtractor; \citealp{bertin:2011}). A reference image was adopted and the roto-translation between the reference image and the $i^{\mathrm{th}}$ image was derived using all sources in the field (except the two components of Luhman\,16). The sample of images was limited to those where the two components of Luhman\,16AB were perfectly resolved and measured. Table\,\ref{tab:summary} indicates the data that were used for the astrometric analysis. The individual images of a single night were stacked together by minimising the scatter in $x$ and $y$ for every source. The resulting uncertainty on the positions, between 6\,mas and 96\,mas for the first season and between 16\,mas and 111\,mas for the second season, was computed as the standard deviation after a 2-$\sigma$ clipping filter. The ICRS ($\alpha$ and $\delta$, i.e.\ the right ascension and the declination, respectively) coordinates were finally restored by measuring the positions of four stars in the first image of each season and deriving a tangent-plane astrometric plate solution. The $\alpha$ and $\delta$ of the four reference stars were extracted from the PPMXL catalog. The accuracy of such a roto-traslation and stacking procedure relies heavily upon a key assumption: no other point source on the field of view is moving over the duration of the observations. Point source here means a single star with a noticeable parallax and/or proper motion or a binary (unresolved or with only one component visible) with a significant orbital motion. At the 10-mas level, no disturbing point source is present in the field of view, thus making the stacked image robust.

\subsection{Astrometric models}
\label{astrometric_models}

Our observational campaign was unfortunately too short to independently derive either the parallax, the proper motion or the orbital motion of the two brown dwarfs in the binary system. We can nevertheless compare our data with the older data in the literature, in particular those from \citet{boffin:2014}, to verify some predictions of the fundamental astrometric parameters and of the relative motion of the two components. The motion of a resolved binary is described by the motion of its barycentre (position, parallax, and proper motion) and the orbital motion of each component around it (Binnendijk 1960). The two orbits only differ by their size (i.e.\ the scale factor $\rho$ in the model equation) and the arguments of periastron, which are $180^{\circ}$ apart.

Furthermore, taking into account two different stacks of reference stars (one per season), we add to the model two linear parameters ($\xi_{\mathrm{off}}$ and $\eta_{\mathrm{off}}$) to cope with some misalignment between the first and the second season dataset. Thirteen parameters are thus required (seven parameters to describe the relative motion of the binary star and five to describe the motion of the barycentre).

The resulting model is therefore given by:
\begin{eqnarray}
\xi^{\prime}&=&\xi^{\prime}_0+\xi_{\mathrm{off}}+f_{\mathrm{a}}\,\pi+\mu_{\xi}(t-2000.0)+BX(t)+GY(t) \nonumber \\
\eta^{\prime}&=&\eta^{\prime}_0+\eta_{\mathrm{off}}+f_{\mathrm{d}}\,\pi+\mu_{\eta}(t-2000.0)+AX(t)+FY(t) \nonumber \
\end{eqnarray}
for the primary,
\begin{eqnarray}
\xi^{\prime\prime}&=&\xi^{\prime\prime}_0+\xi_{\mathrm{off}}+f_{\mathrm{a}}\,\pi+\mu_{\xi}(t-2000.0)+B\rho X(t)+G\rho Y(t) \nonumber \\
\eta^{\prime\prime}&=&\eta^{\prime\prime}_0+\eta_{\mathrm{off}}+f_{\mathrm{d}}\,\pi+\mu_{\eta}(t-2000.0)+A\rho X(t)+F\rho Y(t) \nonumber \
\end{eqnarray}
for the secondary, and
\begin{eqnarray}
\xi&=&\xi_0+\xi_{\mathrm{off}}+f_{\mathrm{a}}\,\pi+\mu_{\xi}(t-2000.0) \nonumber \\
\eta&=&\eta_0+\eta_{\mathrm{off}}+f_{\mathrm{d}}\,\pi+\mu_{\eta}(t-2000.0) \nonumber \
\end{eqnarray}
for the barycentre. $\xi$ and $\eta$ are the standard coordinates\footnote{http://www2.astro.psu.edu/users/rbc/ a501/Girard\_coordinates.pdf}, $t$ denotes the time, $f_{\mathrm{a}}$ and $f_{\mathrm{d}}$ denote the ``\emph{parallax factors}'' (Kovalevsky \& Seidelmann 2004), $\pi$ is the parallax, $\mu_\xi$ and $\mu_\eta$ are the proper motions in $\xi$ and $\eta$, $A$, $B$, $F$, $G$ are the Thiele-Innes parameters \citep{wright:2009}, $\rho$ is the scaling factor of the orbit, $X$ and $Y$ are the so-called elliptical rectangular coordinates (from the solutions of the Kepler motion). In order to compare our findings with the others in the literature, we fit the relative motion over a grid of parameters, because the relative motion is not sufficiently well sampled for performing a real fit in a least squares sense. We discuss the validity of such a simplification in the next section. To focus our analysis on the relative (orbital) motion of the two stars of the binary system we consider the two quantities $\Delta \xi$ and $\Delta \eta$, defined as the differences between the coordinates of the two stars. To compare our findings with others in the literature we modelled $\Delta \xi$ and $\Delta \eta$ with an arc of parabola instead of the true Keplerian orbital model. This assumption works in the limit of a short timespan of the observations with respect to the orbital period. We discuss the validity of such a simplification, for our data and for data in literature, in the next section.

%-------------------------------------
\subsection{Results and discussion}
%-------------------------------------

\begin{figure*}%
\centering
{{\includegraphics[width=5.75cm]{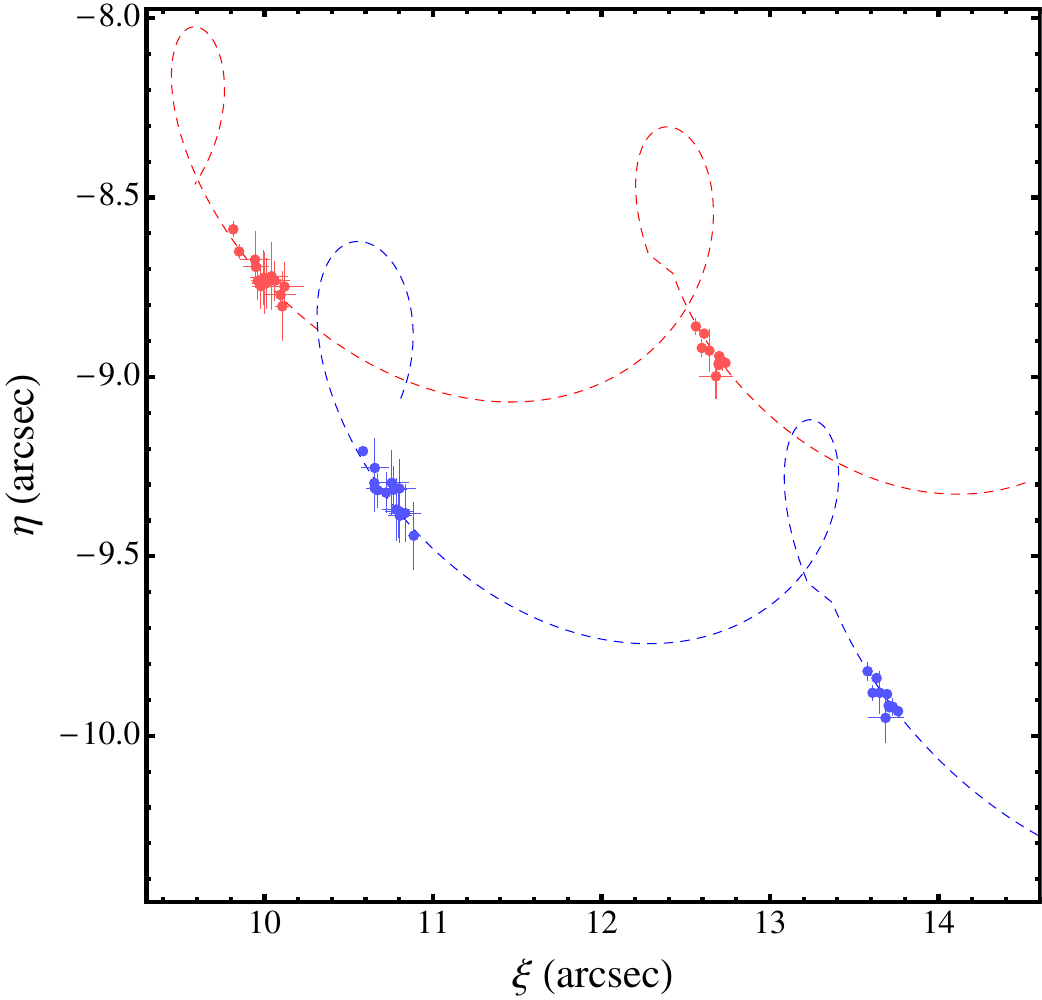} }}%
\qquad
{{\includegraphics[width=5.4cm]{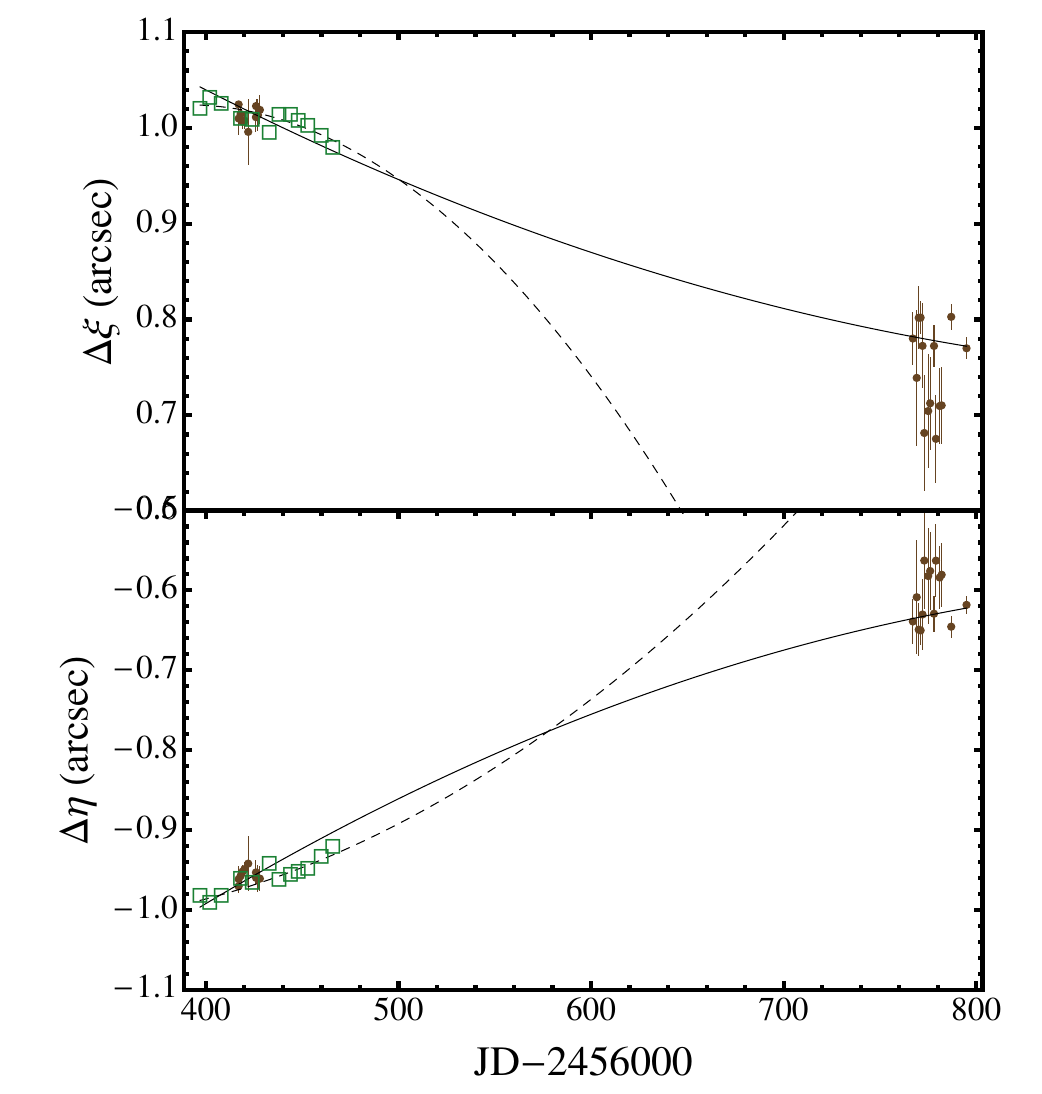} }}%
\qquad
{{\includegraphics[width=5.75cm]{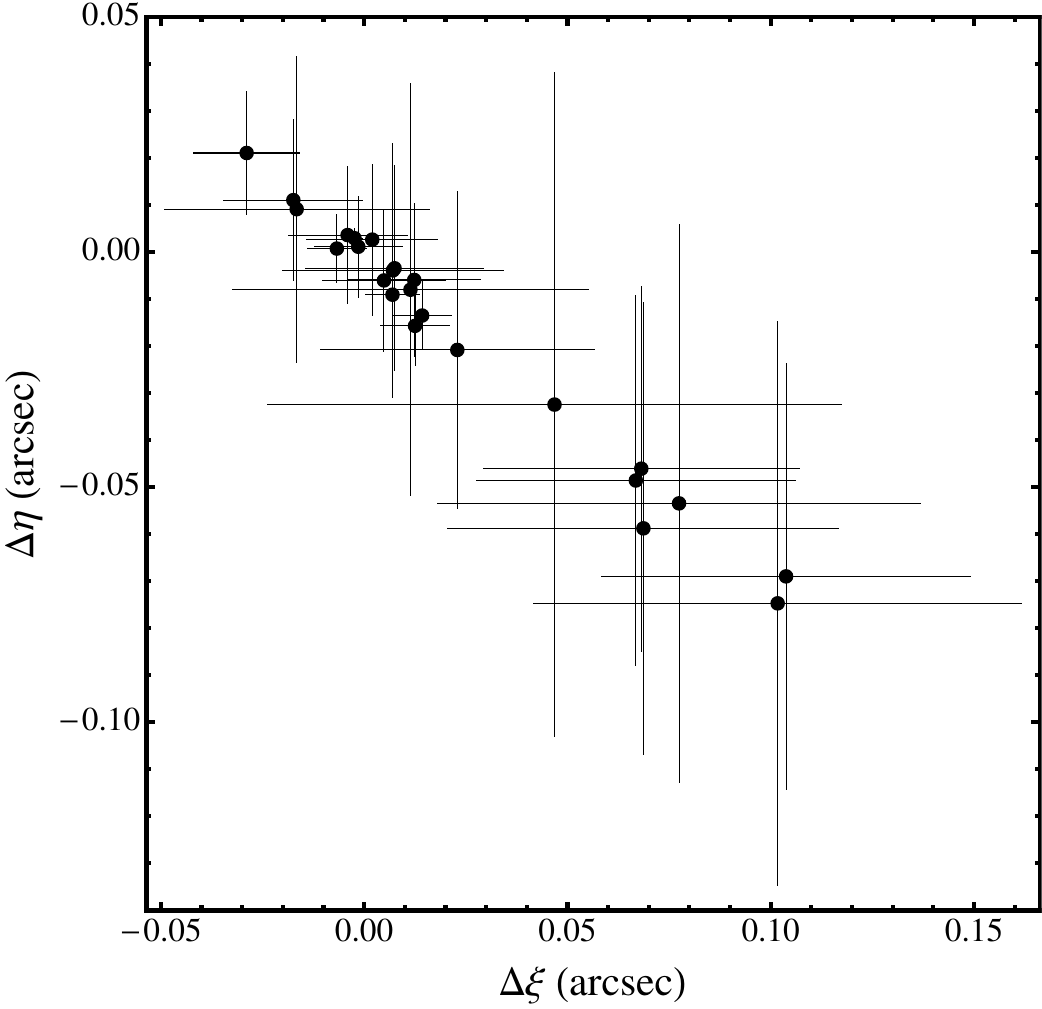} }}%
\caption{{\it Left panel}: absolute positions, $\{\xi,\,\eta\}$, of the components A and B of the binary system
Luhman\,16, estimated from the data taken with the LI camera mounted on the Danish 1.54 m Telescope over two
seasons. Points marked in red (blue) are from Luhman\,16A (16B). Dashed lines represent the best model fits to
the data with the models (for the primary and for the secondary) as described in Sect.\,\ref{astrometric_models}.
{\it Middle panel}: relative positions of the A component with respect to the B component of the binary system
Luhman\,16 as a function of time. Brown points are those estimated with the Danish Telescope over two seasons,
while the open boxes refer to the measurements of \citet{boffin:2014} (the size of their error bars is lower
than the size of the boxes and they were suppressed for clarity). Solid lines represent the best parabolic fits
of both the data sets, while the dashed lines are the fits based on the prediction from \citet{boffin:2014}.
{\it Right panel}: residuals of the LI camera data based on the parabolic fit.}%
\label{position}%
\end{figure*}

%
%\begin{figure}
%\centering
%\includegraphics[width=\columnwidth]{abs_position.eps}
%\caption{
%Absolute positions, $\{\xi,\,\eta\}$, of the components A and
%B of the binary system Luhman\,16, estimated from the data
%taken with the LI camera mounted on the Danish 1.54 m
%Telescope over two seasons. Points marked in blue (red) are from
%Luhman 16A (16B). Dashed lines represent the best fitting of the data
%with the models (for the primary and for the secondary) as described in Sect.\,\ref{astrometric_models}.}%
%\label{abs_position}
%\end{figure}
%
%\begin{figure}
%\centering
%\includegraphics[width=\columnwidth]{rel_position.eps}
%\caption{
%Relative positions of the A component respect to the B component  of the binary system Luhman 16 as a function of time. Brown points are those
%estimated with the Danish Telescope over two seasons, while the
%open boxes refers to the measurements of \citet{boffin:2014} (the size of their error bars are lower than the size of the boxes and were suppressed con clarity).
%Solid lines represent the best parabolic fits of both the data sets, while the dashed lines are the fits based on \citet{boffin:2014} data only.
%}%
%\label{rel_position}
%\end{figure}
%
%\begin{figure}
%\centering
%\includegraphics[width=\columnwidth]{residuals.eps}
%\caption{Residuals of the LI camera data (see Fig.\,\ref{rel_position}) based on the parabolic fit.
%}%
%\label{residuals}
%\end{figure}
%
The results of the least-squares fit of the absolute positions of the components A and B (i.e.\ the model from Sect.\,\ref{astrometric_models}) are plotted in the left panel of Fig.\,\ref{position}. In the least-squares fit we kept the proper motion and parallax values fixed to those of \citet{boffin:2014}. So we really ``adjust'' the constants, the offsets and the orbital solution (over a grid of parameters). We fit simultaneously the coordinates of the two stars. The accuracy of such a fitting procedure relies upon an assumption: if any systematics are present in our data, they affect evenly the two components. This is reasonable taking into account the short distance ($\sim 1$\,arcsec) of the two components. The consequence of such an assumption is a symmetry in the $\chi^{2}$ calculated, using the simultaneous best fit model, over the data set of each component (left panel of Fig.\,\ref{position}). This symmetry seems to be violated. The contribution of the primary to the $\chi^{2}$ is $\sim 18\%$ larger than that due to the secondary. Under our hypothesis of no-graded systematic effects, we suggest three possible scenarios:
\begin{itemize}
\item the orbital solution of the binary system is not accurate;
\item the proper motion estimate is not accurate;
\item a companion might be present around component A, making it oscillate.
\end{itemize}
We compared the parabolic least-squares fit of the relative positions with the results of \citet{boffin:2014}. Both $\Delta \alpha$ and $\Delta \delta$ are fitted with distinct parabolae (solid lines in the middle panel of Fig.\,\ref{position}). It appears clear that the curvature predicted by the parabolic fit over the two months of data (dashed lines) by \citet{boffin:2014} is not supported by our findings after one year. This is not unexpected because \citet{boffin:2014} estimated the probability of rejecting the parabola by accident to be 12.95\%. As shown by \citet{boffin:2014}, the residuals of the parabolic fit are highly correlated (right panel of Fig.\,\ref{position}). This correlation in our data does not have the same amplitude as the one presented in \citet{boffin:2014} and it seems to be correlated with the amplitude of the errors. This suggests to interpret the correlation as a systematic effect due to the roto-translation procedures. Actually, our precision does not allow us to confirm or reject the presence of a planetary signature but our findings highlight the need for an improved determination of the orbit of the binary system.
%Therefore, a new observational campaigns is highly recommended.

% Sect. 4
%%%%%%%%%%%%%%%%%%%%%%%%%%%%%%%%%%%%%%%%%%%%%%%%%%%%%%
\section{Summary}
\label{summary}
%%%%%%%%%%%%%%%%%%%%%%%%%%%%%%%%%%%%%%%%%%%%%%%%%%%%%%

We have photometrically monitored the brown-dwarf binary system Luhman\,16AB over two seasons, in 2013 and 2014, for a total of 38 nights. For this task, we have utilised the EMCCD LI camera on the Danish 1.54\,m Telescope and the target was observed in a wavelength range corresponding to a combination of the SDSS $i$ and $z$ filters. Thanks to this instrumentation, we were able to collect 1132 high-resolution images, in which the two components of the binary system are consistently well-resolved.

Between April and May 2014, we obtained an optimum time coverage during sixteen nights, fourteen of them continuous. The data from the 16-night time series were used to analyse the variability of the two brown dwarfs, which is likely caused by the circulation of clouds in their atmospheres, and to estimate their rotational velocity. For this purpose, we fitted the photometric data for each component with a Gaussian process model. In the case of Luhman\,16A, the hotter component, we estimated that the most probable rotation period is $\sim 8$\,hr, suggesting that it rotates more slowly than its companion and their rotational axes are therefore well aligned. For the colder component, Luhman\,16B, we estimated $5.1 \pm 0.1$\,hr for its rotation period, which is in a very good agreement with previous estimates \citep{gillon:2013,burgasser:2014}.

Data from both 2013 and 2014 were used for a detailed astrometric analysis and for investigating the possible presence of an additional small companion in the system, as proposed by \citet{boffin:2014}. Our two-season monitoring is not consistent with the predicted motion of Luhman\,16AB by \citet{boffin:2014}, which was based only on a two-month dataset. However, our data have insufficient phase coverage and precision to perform a conclusive analysis of the parallax, proper motion and relative motion of the binary system. Ultimately, we cannot confirm or reject the presence of any astrometric signal induced by a massive planet or low-mass brown dwarf as hinted by \citet{boffin:2014}. Further high-cadence astrometric monitoring of the Luhman\,16AB system is thus highly encouraged, particularly to provide a set of measurements coincident with those that are being collected by Gaia, which is expected to deliver astrometry for the system at the milli-arcsecond level (e.g., \citealp{sozzetti:2014}).

%  Acknowledgements
%%%%%%%%%%%%%%%%%%%%%%%%%%%%%%%%%%%%%%%%%%%%%%%%%%%%%%

\begin{acknowledgements}
The operation of the Danish 1.54\,m telescope is financed by a grant to UGJ from the Danish Natural Science Research Council (FNU). The reduced light curves presented in this work will be made available at the CDS. J\,Southworth acknowledges financial support from STFC in the form of an Advanced Fellowship. OW and J\,Surdej acknowledge support from the Communaut\'{e} fran\c{c}aise de Belgique - Actions de recherche concert\'{e}es - Acad\'{e}mie Wallonie-Europe. SHG and XBW would like to thank the financial support from National Natural Science Foundation of China through grants Nos.\ 10873031 and 11473066. MH acknowledges support from the Villum Foundation. We acknowledge the use of the following internet-based resources: the ESO Digitized Sky Survey; the TEPCat catalogue; the SIMBAD data base operated at CDS, Strasbourg, France; and the arXiv scientific paper preprint service operated by Cornell University.
\end{acknowledgements}

\bibliographystyle{aa}

%%%%%%%%%%%%%%%%%%%%%%%%%%%%%%%%%%%%%%%%%%%%%%%%%%%%%%
%

\end{document}